\documentclass{emulateapj}

\newcommand{\kms}{\,km~s$^{-1}$}   \newcommand{\sqcm}{\,cm$^{-2}$}  
\newcommand{\os}{\ion{O}{6}}       
\newcommand{\hi}{\ion{H}{1}}       \newcommand{\cf}{\ion{C}{4}}
\newcommand{\lya}{Lyman-$\alpha$}  
     \newcommand{\hst}{\emph{HST}}
\newcommand{\tm}{\tablenotemark}   \newcommand{\tn}{\tablenotetext}
\newcommand{\nf}{\ion{N}{5}}       \newcommand{\fuse}{\emph{FUSE}}
\newcommand{\ntot}{775}            \newcommand{\hw}{\ion{H}{2}}
\newcommand{\ngal}{447}            \newcommand{\nigm}{328}

\begin{document}
\shorttitle{\os\ across Cosmic Time}
\shortauthors{Fox}
\title{The surprisingly constant strength of \os\ absorbers over cosmic time}
\author{Andrew J. Fox}
\affil{European Southern Observatory, Alonso de C\'ordova 3107, Casilla
  19001, Vitacura, Santiago, Chile}
\affil{Institute of Astronomy, University of Cambridge, Madingley
Road, Cambridge, CB3 0HA, UK}
\email{afox@eso.org}
%\received{October 25, 2010}
%\accepted{January 23, 2011}

\begin{abstract}
\os\ absorption is observed in a wide range of astrophysical
environments, including the Local ISM, the disk and halo of the Milky Way, 
high-velocity clouds, the Magellanic clouds, starburst galaxies, 
the intergalactic medium, damped \lya\ systems, and gamma-ray-burst
host galaxies. Here a new compilation of \ntot\ \os\ absorbers drawn
from the literature is presented, all observed at high resolution
(instrumental FWHM$\le$20\kms) and covering the redshift range $z$=0--3.
In galactic environments [log\,$N$(\hi)$\ga$20],
the mean \os\ column density is shown to be insensitive to
metallicity, taking a value log\,$N$(\os)$\approx$14.5 for
galaxies covering the range --1.6$\la$[O/H]$\la$0.
In intergalactic environments [log\,$N$(\hi)$<$17], the mean \os\
component column density measured in datasets of similar sensitivity
shows only weak evolution between $z$=0.2 and $z$=2.3, but IGM \os\ components
are on average twice as broad at $z$=0.2 than at $z$=2.3. 
The implications of these results on the origin of \os\ are discussed.
The existence of a characteristic value of log\,$N$(\os) for 
galactic \os\ absorbers, and the lack of evolution in log\,$N$(\os)
for intergalactic absorbers,
lend support to the ``cooling-flow'' model of \citet{He02}, in which
all \os\ absorbers are created in regions of initially-hot
shock-heated plasma that are radiatively cooling through coronal
temperatures. These regions could take several forms, including 
conductive, turbulent, or shocked boundary layers between warm
($\sim$10$^4$\,K) clouds and hot ($\sim$10$^6$\,K) plasma,
although many such layers would have to be intersected by a typical
galaxy-halo sightline to build up the characteristic galactic $N$(\os). 
The alternative, widely-used model of single-phase photoionization for
intergalactic \os\ is ruled out by kinematic evidence in the majority
of IGM \os\ components at low and high redshift.
\end{abstract}


\section{Introduction}
In a ground-breaking paper discussing the theoretical basis for the hot
Galactic corona, \citet{Sp56} correctly predicted 
that the lithium-like O$^{+5}$ ion ``might be sufficiently abundant to
produce measurable absorption'' in the UV spectra of background sources, 
and could therefore be used to detect and analyze such a corona. 
\os\ absorption in its far-ultraviolet resonance doublet at
1031.926 and 1037.617\,\AA\ 
is now routinely detected in the interstellar medium (ISM) of the Milky
Way, the extended halos of other galaxies, and the
intergalactic medium (IGM) from $z$=0 to $\approx$3.
The high cosmic abundance of oxygen, the intrinsic strength of the
\os\ doublet, and the ability to probe warm/hot gas all combine to
make \os\ a powerful and well-studied tracer of the diffuse Universe.

It is common practice to study the detailed properties of \os\
absorption in a specific interstellar or intergalactic location. An
alternative, global approach is to compare \os\ measurements from many
different locations, and search for correlations with other observable
parameters (such as redshift and metallicity). Along these lines,
\citet[][hereafter H02]{He02} synthesized \os\ observations across
several low-redshift 
environments, and proposed a unified model in which all \os\ absorbers
(both galactic and intergalactic) trace radiatively-cooling regions of
initially-hot gas. In the last eight years, many new high-resolution
(instrumental FWHM$\le$20\kms) \os\ datasets have become available. 
In order to draw them together, and to gain insight into the nature
of \os\ absorbers, a new, heterogeneous
compilation of \ntot\ \os\ absorbers is presented in this paper. 
The compilation includes galactic \os\ covering 
galaxy metallicities in the range $\approx$0.01--1 times solar, and
intergalactic \os\ covering redshifts from $z$=0 to $z\!\approx\!2.5$.
In \S2 the relevant \os\ ionization physics is briefly reviewed.
In \S3, a survey of all published interstellar and
intergalactic \os\ absorbers is presented.
The new galactic and intergalactic compilations
are presented in \S4 and \S5, respectively,
together with a discussion of their properties. The key results are
summarized in \S6.

\section{Brief overview of \os\ ionization physics}
The energy required to ionize O$^{+4}$ to O$^{+5}$
(the ion traced by \os) is 113.9\,eV (8.4 Rydbergs), whereas ionizing
O$^{+5}$ to O$^{+6}$ requires 138.1\,eV (10.2 Rydbergs). 
In a given plasma, the \os\ ionization fraction $f$(\os)$\equiv$\os/O
will be set by the balance between ionization from lower states and
recombination from higher states.
The ionization process can be either ion-electron collisions
in high-temperature plasma (collisional ionization)
or photoionization by extreme-UV radiation.
Collisional ionization of \os\ requires ``coronal'' plasma 
temperatures of $\sim$10$^{5-6}$\,K, with the ion fraction $f$(\os)
reaching a maximum of $\approx$0.22 near 300\,000\,K \citep{SD93, Br06, GS07} 
in the case of collisional ionization equilibrium (CIE). However,
non-equilibrium conditions are likely
given that the interstellar cooling function peaks at coronal
temperatures, and that coronal plasma is
thermally unstable \citep{Ka73, SM76, EC86, Wi09}. Under non-equilibrium
conditions, the cooling time can be shorter than the recombination
time, resulting in ``frozen-in'' \os\ persisting to lower temperatures
than would hold under CIE.
This has the important consequence that narrow \os\ components are
not necessarily photoionized; they can also trace cooled,
initially-hot plasma. Given the additional complication of
non-thermal broadening, \os\ line-widths ($b$-values) by themselves do not
offer a clean discriminator between ionization mechanisms.
Photoionization of \os\ is possible if the density of EUV photons with
$E\!>\!113.9$\,eV is high enough, e.g. near AGN and gamma-ray bursts (GRBs).
In the Galactic ISM (and by extension, in the ISM of external
galaxies), the ionizing radiation field has a sharp break at 54\,eV
caused by the \ion{He}{2} edge in hot-star spectra \citep{BH86}, 
and so photoionization is ruled out as the origin of the observed \os\
absorption.  
In the IGM, photoionization models \citep{Ha97, Be02, Le06, Ho09}
indicate that the production of \os\ by the extragalactic background
(EGB) radiation requires an ionization parameter 
$U\!\equiv\!n_\gamma/n_{\rm H}$ 
(the ratio of the ionizing photon density to gas density)
in the range $\approx$0.1--1 at any redshift, so long as the shape of
the EGB does not change substantially, which is reasonable for the
redshift range $z$=0--3 over which \os\ can be observed \citep{HM96, HM01}.

\section{Published \os\ Absorption Detections}

In this paper, \emph{galactic} (interstellar) and \emph{intergalactic} \os\
absorbers are distinguished by a simple cut made in \hi\
column density, referring to absorbers with $N$(\hi)$>$10$^{17}$\sqcm\ and
$N$(\hi)$<$10$^{17}$\sqcm, respectively.
This approach is powerful since the \hi\ lines can often be measured
in the same spectra as the \os\ lines.
Of course, the galactic/intergalactic division is not clear-cut and
somewhat arbitrary; \citet{WS09} report that \emph{all} low-$z$ \os\
absorbers arise within 550\,kpc of an $L\!>\!0.25L_\star$ galaxy, so all \os\
absorbers could be called galactic at some level \citep[see also][]{St06}. 
Nonetheless, the distinction at 10$^{17}$\sqcm\ is still useful: it
represents the transition where an absorber becomes optically thick to
hydrogen-ionizing radiation at $\lambda\!<\!912$\,\AA. 
The distinction also fits historically-defined observational categories: 
intergalactic absorbers are traced by the \lya\ forest, whereas
galactic absorbers can be divided into Lyman Limit Systems (LLSs) with
17$<$log\,$N$(\hi)$<$19, sub-damped \lya\ systems (sub-DLAs or super
LLSs) with 19$<$log\,$N$(\hi)$<$20.3, and genuine damped \lya\ systems (DLAs)
with log\,$N$(\hi)$>$20.3, which represent structures of progressively
higher overdensity.

In order to prepare the compilations, a review of all published 
detections of interstellar and intergalactic \os\ absorption is now
presented. Readers interested in the new results drawn from the
compilations may wish to skip to \S4.  

\subsection{Galactic (Milky Way) \os}
The first detections of interstellar \os\ absorption
were made with the {\it Copernicus} satellite, 
along sightlines through the Galactic disk 
\citep{Ro73, Yo74, Yo77, JM74, Je78a, Je78b, Co79, SC94}.
This revealed a network of highly-ionized interstellar clouds
seen in the form of low-velocity ($|v_{\rm LSR}|\!\la\!100$\kms)
\os\ absorption components.
A small number of \os\ detections were made with the Astro-1 \citep{Da93} and
ORFEUS Space Shuttle missions \citep{Hu98, Wi98, Se99},
then the study of interstellar \os\ absorption flourished following the
launch of the {\it Far-Ultraviolet Spectroscopic Explorer 
(FUSE)} satellite in 1999. Many \fuse\ studies of \os\ took the form of
surveys over many lines-of-sight, each targeting
different regions of the Galaxy,
either the local ISM \citep{Oe05, SL06, WL08, Ba10},
the Galactic disk \citep[latitudes $|b|\!<\!10\degr$;][]{Bo08, Le11}, 
or the Galactic halo 
\citep[$|b|\!>\!10\degr$;][]{Sa00, Sa03, Wa03, Zs03, IS04b, SW09}.
Other \fuse\ studies analyzed the \os\ absorption
in individual disk and halo directions 
\citep{Ri01a, Ni02, St02, Fo03, Ho03, We04, Wl05, Wl06, Wl07, Ya09a},
and the small-scale structure in \os\ using closely-spaced
(degree-scale) sightlines \citep{Ho02b}.
{\it Chandra} X-ray observations of zero-redshift (interstellar)
absorption in the K-shell line of \os\ at 22.040\AA\ have also been
reported \citep{Ni05, Wl05, Ya09b}.

In addition to these low-velocity detections, \os\ is also commonly
detected in high-velocity clouds (HVCs), defined as absorption features
at $|v_{\rm LSR}|\!>\!100$\kms\ (and in practice $|v_{\rm LSR}|\la400$\kms)
seen in the spectra of background AGN
\citep{Se00, Se03, Mu00, Ri01b, Tr03, Ni03, Co03,Co04,Co05,Co07,
  Fo04,Fo05,Fo06, Ga05, Ke06}
or distant halo stars \citep{Ze08}.
HVCs trace a variety of processes in the extended gaseous halo of the
Milky Way, including accretion, outflow, and the stripping of gas from
nearby satellites.
\os\ absorption has also been detected in individual Galactic
interstellar objects, such as the Cygnus Loop supernova remnant
\citep{Bl02, Bl09} and the Southern Coalsack dark nebula \citep{An04}.

\subsection{\os\ in nearby galaxies}
\os\ absorption is detected in the LMC
\citep{Fr00, Bl00, Ho02a, Da02, DB06, LH07, Le09b, Pa10}, the SMC 
\citep{Hp02, Da02}, and the Magellanic Bridge \citep{Le02} and
Magellanic Stream \citep{Se03, Fo05, Fo10}, which trace the
interactions between the two Magellanic Clouds and the Milky Way.
The typical \os\ column densities in the LMC \citep{Ho02a, Pa10} are
comparable to those in the Milky Way halo; those seen in the SMC
\citep{Hp02} are slightly higher.
Strong \os\ absorption is detected in the starburst galaxy
NGC~1705 \citep{He01}, the merging galaxy VV~114 \citep{Gr06},
the UV-luminous galaxy Haro~11 \citep{Gr07},
and in 12 of the 16 low-redshift starburst galaxies studied by \citet{Gr09}.

\subsection{Low-redshift ($z\!\la\!0.5$) intergalactic \os}
The study of low-redshift intergalactic \os\ absorption was enabled by
the launch of \hst\ and the installation of a series of
high-resolution ultraviolet spectrographs (GHRS, STIS, and COS). 
Following the detection of \os\ at $z$=0.14232 toward QSO PG~0953+415 by
\citet{Tr00}, detections and detailed analyses of
low-redshift intergalactic \os\ absorbers were reported
by many groups. Most of these studies focused on individual sightlines 
\citep{TS00, Tr01, Tr06, Se01, Se04, Sa02, Sa05a, Sa05b, Wi06, Wi10, 
  Le06, Le09a, Na10b, Na10c, Oe00, Je03, Ri04, Pr04, Tu05,
  Co08, Ho09, LC10, Da10}. Others took the form of surveys of all
available data \citep{DS05, DS08, Da06, Tr08, TC08a, TC08b}.
\citet{Br04} report three \os\ absorption-line detections
in sight lines passing through low-$z$ galaxy clusters, arguing that
these detections trace intergalactic filaments.

The association between intergalactic \os\ absorbers
and nearby galaxies has been investigated by
searching the fields around bright quasars
for nearby galaxies \citep{St06, Pr06, WS09, CM09}, and correlating the
absorber redshifts with the galaxy redshifts. 
\citet{WS09} report that all \os\ absorbers arise within 550\,kpc of
an $L\!>\!0.25L_\star$ galaxy, and there is no evidence for \os\ in
intergalactic voids \citep{St06, St07}, defined as absorbers at 
$>$1.4\,Mpc from the nearest $L\!>\!L_\star$ galaxy.
Other galaxy/\os\ relationships have been investigated 
via the \os\ detections (i) at $z$=0.225 and $z$=0.297 toward QSO~H1821+643
\citep{Sa98}, close in redshift to two nearby spiral galaxies,
(ii) in the Lyman Limit System at $z$=0.16716 toward QSO~PKS~0405-123,
thought to be associated with a pair of nearby galaxies \citep{CP00, Sa10},
and (iii) in the halo of a $\approx\!2L_\star$ spiral galaxy at $z$=0.22
\citep{Na10a}. 

An important subset of the IGM absorbers are the ``proximate''
(also called ``associated'' or ``$z_{\rm abs}\!\approx\!z_{\rm em}$'')
absorbers, defined as those within a fixed velocity interval
(typically 5\,000\kms, but sometimes 3\,000, 2\,500\kms\ or 1\,500\kms)
of the background QSO \citep[e.g.][]{Fo86, Tr08, Fo08a}. 
Such proximate absorbers are usually excluded from intervening samples
in order to avoid truly intrinsic absorbers, proximity
effects, and environmental effects.
Truly intrinsic absorbers, tracing inflowing and outflowing gas arising
near the central engines of AGN, represent a separate
category of object and lie outside the scope of this paper. 
Fortunately, they can usually be identified by their saturated
high-ion profiles, time variation of absorption, evidence for partial
coverage of the continuum source, and/or presence of excited-state
absorption lines \citep[e.g.][and references therein]{HF99, SH10}. 
The remaining (non-intrinsic) proximate absorbers are included in the
compilation presented in this paper.

\subsection{Intermediate-redshift ($0.5\!\la\!z\!\la\!2.0$) intergalactic \os}
Intermediate-redshift intergalactic \os\ absorption was first
detected with the low-resolution \hst/FOS spectrograph 
by the surveys of \citet{Be94} and \citet{Ja98}. Detailed studies
of individual intermediate-$z$ absorbers followed, both with FOS
\citep{CC99, Lo99} and STIS/E230M  
\citep{Re01,Re06, Lv02,Lv03a, Mi07}.
Most of these studies have probed the range 1.0$\la\!z\!\la$2.0; the
range 0.5$\la\!z\!\la$1.0 remains largely unobserved in \os,
largely because of the small effective
area of near-UV space-based spectrographs, and the shortage of
UV-bright AGN at sufficiently high redshift.

\subsection{High-redshift ($z\!\ga\!2.0$) intergalactic \os}
Above $z\!\approx\!2.0$ the \os\ doublet becomes detectable
from the ground, where it can be measured using high-resolution
spectrographs. However, high-redshift \os\ detections are complicated
by the possibility of blending with intervening \lya\ forest
absorbers. The rapidly increasing density of the forest with
increasing redshift makes \os\ absorption detections in individual
systems above $z\!\approx\!3$ very challenging, but not impossible
\citep[][]{BT96, Fr10a}.  

The earliest evidence for the presence of intergalactic \os\ absorption
at $z\!\ga\!2.0$ came from features seen in composite spectra
formed by stacking individual frames \citep{LS93, Da98}. 
Since then, high-$z$ IGM \os\ detections have been reported in many
individual systems studied at high resolution
\citep{Be02,Ca02,Si04,Si06, Lv03b,Lv04,Lv09, Ag05,Ag07, Lo07,Sy07,Go08,FR09}, 
and several high-$z$ \os\ surveys exist \citep{Si02, BH05, Fo08a}.
Furthermore, 1756 \os\ doublet candidates (of which 145 fall in the
most secure category) have been reported in a SDSS
database of 3702 QSOs observed at low resolution \citep{Fr10a, Fr10b}.
A complementary, statistical approach for detecting intergalactic \os\
is the pixel optical depth method, where one searches pixel-by-pixel for 
a \lya\ signal and measures whether absorption appears at the
corresponding \os\ pixel \citep{Sy00, Ar04, PH04, Pi06, Pi10, Ag08}. 
This method is increasingly being used on large samples of
low-resolution spectra, and allows \os\ to be detected out to $z>3$
\citep{Pi10}. 

\subsection{High-redshift ($z\ga2.0$) galactic \os}
\os\ detections have been reported in each of the three classes of
optically-thick (galactic) quasar absorption-line system at $z\!\ga\!2.0$: 
in LLSs \citep{KT97, KT99, DO01}, in sub-DLAs \citep{Si02, Fo07b}, and in DLAs
\citep{Fo07a, Le08, El10}. A composite spectrum of 341 DLAs in the SDSS
database shows clear \os\ absorption \citep{Ra10}. 
\os\ has also been detected in the host
galaxies of high-redshift GRBs \citep{DE07, DE10, Fo08b}, 
which can be viewed as a subset of DLAs, known as GRB-DLAs, though 
in these cases there may be a contribution to the \os\ from the
circumburst region immediately surrounding the GRB \citep{Pr08}.

\section{Compilation of galactic \os}
In Figure 1 the new compilation of galactic \os\ absorbers is presented.
It includes a total of \ngal\ \os\ measurements made at high
resolution, covering %*
24 Local ISM sightlines \citep[LISM;][]{SL06},
131 Galactic disk sightlines \citep{Bo08}, 
91 Galactic halo sightlines \citep{Sa03}, 
84 HVCs \citep{Se03}, 
18 SMC sightlines \citep{Hp02}, 
70 LMC sightlines \citep{Pa10}, 
12 starburst galaxies with \os\ detections \citep{Gr09}, 
12 DLAs at $z$=2--3 \citep{Fo07a},
and 5 GRB-DLAs at $z$=2--4 \citep{Fo08b}. 
All these environments have log\,$N$(\hi)$\ga$17, and most have 
log\,$N$(\hi)$\ga$20.
The galactic compilation is summarized in Table 1,
with details of each set of measurements given in the footnotes.
All \os\ measurements (except for the Galactic disk, where profile
fitting was used) were made using the apparent optical depth 
(AOD) method \citep{SS91}, and were integrated over the observed range of
absorption. They were retrieved from tables in the online versions of
the papers cited (except the GRB-DLA AOD measurements, which were
remade by the author). 
The AOD method gives accurate measurements of the true column density
unless the lines are saturated, in which case the results are lower limits.
Saturation is present for all \os\ absorbers in the GRB-DLA sample,
and for several objects in the starburst galaxy sample.

\begin{figure*}[!ht]
\epsscale{1.12}
\plotone{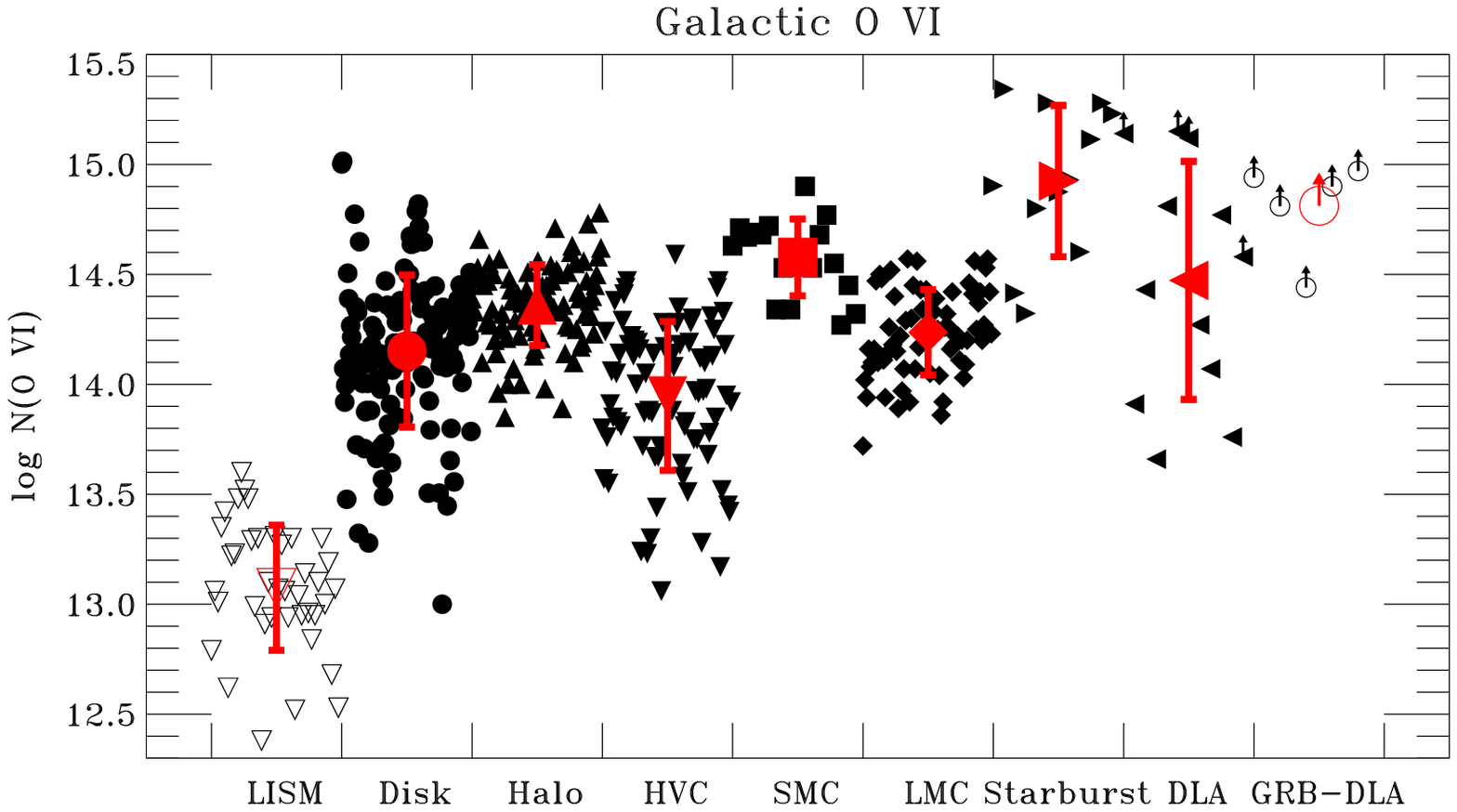}
\caption{Comparison of \os\ column density in various \emph{galactic}
 environments, derived from apparent optical depth
 measurements unless noted otherwise. The following
 samples are shown, ordered by increasing distance:
 Local ISM \citep{SL06}, Galactic disk \citep[][profile-fit
 measurements]{Bo08}, Galactic halo 
 \citep{Sa03}, HVCs \citep{Se03}, the SMC \citep{Hp02}, the LMC
 \citep{Pa10}, starburst galaxies \citep{Gr09}, 
 DLAs at $z$=2--3  \citep{Fo07a},
 and GRB-DLAs (GRB host galaxies) at $z$=2--4 \citep{Fo08b}, with a
 different symbol used for each sample.
 Within each bin, the data points are spread in the x-direction for
 clarity. Red symbols with error bars show the mean and standard
 deviation of log\,$N$(\os) in each category. For details of each
 sample see Table 1. Note that the LISM and HVC measurements are made
 over partial (rather than extended) galaxy-halo sightlines.}
\end{figure*}

\begin{figure}[!ht]
\epsscale{1.12}
\plotone{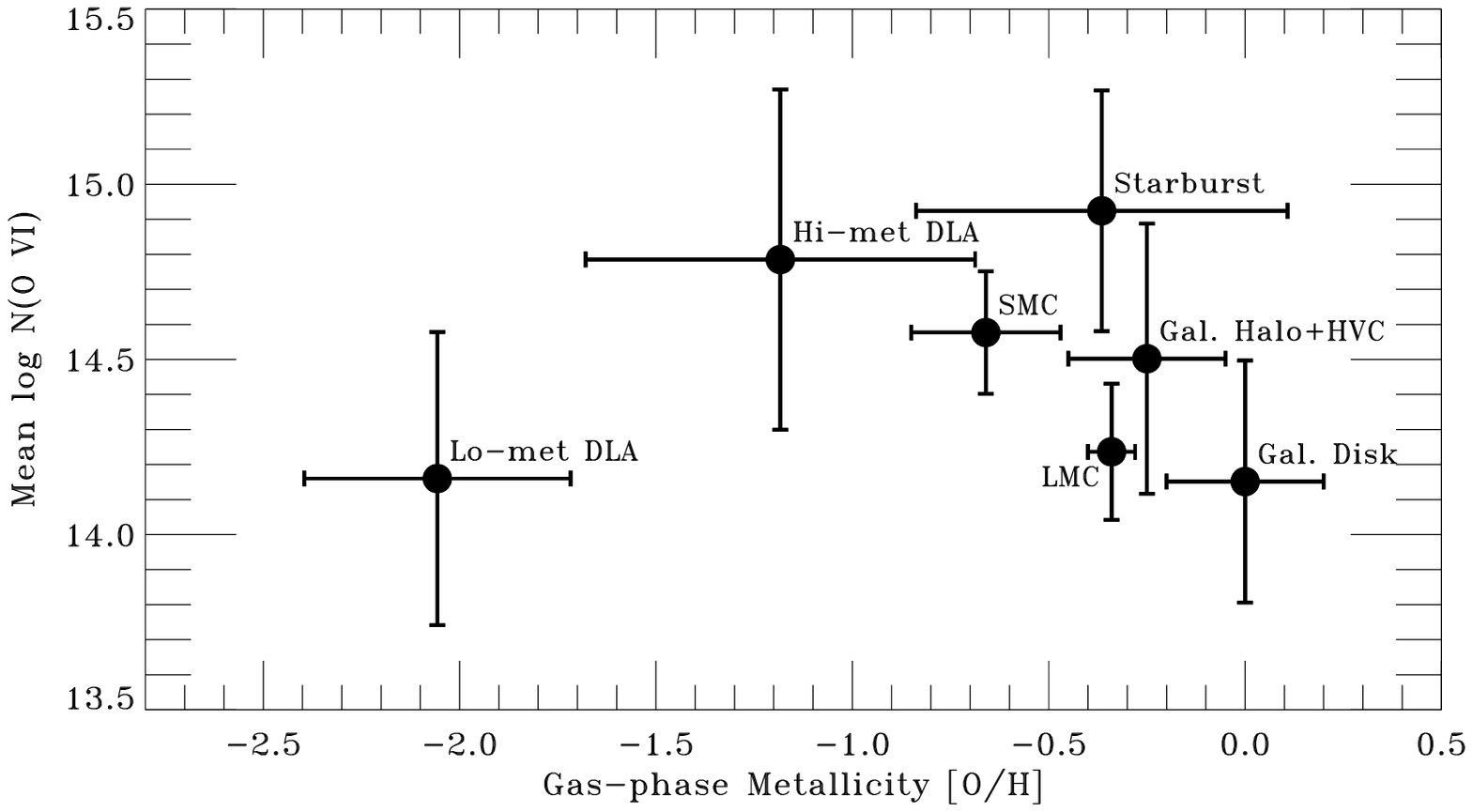}
\caption{Mean \os\ column density versus gas-phase metallicity for 
  seven samples of galactic \os\ absorbers:
  the Galactic disk \citep{Bo08}, Galactic halo plus HVCs \citep{Sa03, Se03},
  the SMC \citep{Hp02}, the LMC \citep{Pa10}, 
  starburst galaxies \citep{Gr09}, and DLAs at $z$=2--3 
  split into low-metallicity and high-metallicity sub-samples \citep{Fo07a}. 
  Considering the range of metallicities (and masses) probed,
  $\langle N$(\os)$\rangle$ is surprisingly insensitive to metallicity.}
\end{figure}

On the left side of Figure 1 are the LISM \os\ measurements,
made along short ($\sim$40--200\,pc) sightlines 
with a well-known geometry \citep{SL06}.
The \os\ in the LISM is thought to arise at the interface
between the Local Bubble, the cavity of hot ($\sim$10$^6$\,K) gas immediately
surrounding the Sun, and the neutral gas beyond \citep{SL06, Ba10}.
This is supported by the observation that the typical \os\ column density
measured along LISM sightlines is close to the value predicted
theoretically in a single conductive interface,
$N$(\os)$\approx$10$^{13}$\sqcm\ \citep{Bo90, GS10}.
The LISM column densities cannot be easily compared to those measured
in extended sightlines through other galactic halos, which cover
length scales on the order of kpc to tens-of-kpc.
All the extended-sightline samples shown on Figure 1
(i.e. all samples except the LISM and HVCs) 
have mean \os\ columns confined to a
narrow range between log\,$N$(\os)=14.1 and 14.9 (the Galactic disk
sight lines studied by \citet{Bo08} have lengths of
$\approx$0.8--10\,kpc, and are therefore treated here as extended).  

To emphasize the relatively small range in \os\ column shown by these
diverse galaxies, in Figure 2 the mean log\,$N$(\os) is plotted versus
gas-phase metallicity for the extended-sightline samples.
Here the mean \os\ column densities through
the Milky Way and HVCs have been summed to determine the total \os\
column in a (one-sided) sightline through the Milky Way halo
(integrated over velocity), allowing a better comparison with the other
galactic environments. The main finding of Figure 2 is that 
{\it the mean column density for galactic \os\ is surprisingly
constant over two orders of magnitude of metallicity, with the
Galactic disk, the Galactic halo, the LMC, the SMC, starburst
galaxies, and DLAs at $z$=2--3 all showing 
$\langle$log\,$N$(\os)$\rangle$ within 0.4\,dex of 14.5}. 
The existence of a characteristic log\,$N$(\os) in galactic
environments has been noticed
before \citep[H02,][]{Sa03}. What is new here is the observational
finding that the insensitivity of $N$(\os) to galaxy metallicity
extends down to $-$1.6$\la$[O/H]$\la$--0.6, to the upper end of the
DLA metallicity regime. Below $-$1.6, $\langle$log\,$N$(\os)$\rangle$
falls off and is no longer independent of metallicity: the
low-metallicity sub-sample of DLAs shows a mean log\,$N$(\os)
0.5\,dex lower than the high-metallicity sub-sample \citep{Fo07a}.
If one reduces the mean DLA \os\ column by a factor of two
(0.3 dex) to correct for the two-sided nature of a DLA-halo sightline
(as opposed to the Milky Way, LMC, SMC, and
starburst galaxy cases, where the sightline only passes through one
side of the halo), then the constancy of \os\ (down to [O/H]$\approx$--1.5)
becomes even more striking.

The non-dependence of $N$(\os) on metallicity has an important consequence
on the total ionized hydrogen column in the \os\ absorbers
$N$(\hw)$_{\rm O~VI}$, which can be written as 
$N$(\hw)$_{\rm O~VI}$=$N$(\os)/[(O/H)$f$(\os)], 
and so scales inversely with metallicity.
If the ionization fraction $f$(\os) is treated as a constant
(as is possible if a common origin mechanism applies),
then the constant value of $N$(\os)
implies that the lower-metallicity galaxies contain
\emph{larger} columns of ionized hydrogen at \os-bearing temperatures.
For example, a typical sightline through a high-metallicity DLA, with  
$\langle N$(\os)$\rangle$=14.79 and $\langle$[O/H]$\rangle$=$-$1.2,
contains an average $N$(\hw)$_{\rm O~VI}$ a factor of $\approx$17
higher than a typical sightline through the Milky Way halo, which has
$\langle N$(\os)$\rangle$=14.50 and an assumed [O/H]$\approx-0.25$.

\subsection{Discussion on Origin of Galactic \os}
Detailed studies of the high ions in the Milky Way (the galaxy with
the best-studied gaseous halo) have emphasized the complexity of
the halo environment, with many astrophysical
processes potentially at work \citep{Sa03, IS04b, Bo08},
and there is no obvious reason for other galactic halos to be any simpler.
Therefore the existence of a characteristic value for galactic $N$(\os) is
somewhat surprising, and models that can explain it deserve
attention. Such a model was suggested by 
H02, who argued that the insensitivity of $N$(\os) 
to metallicity could be explained if
the \os\ arises in radiatively-cooling regions of
initially-hot shock-heated plasma that are passing through the
coronal regime \citep[see also][]{Fu05}. In such a
``cooling flow'', $N$(\os)=3$kTv_{\rm cool}{\rm (O/H)}_\odot
Zf$(\os)$/\Lambda$, where $v_{\rm cool}$ is the cooling-flow velocity
and $\Lambda$ the cooling function (H02). Since $\Lambda$ is almost linearly
proportional to $Z$, the $Z$ terms cancel out, elegantly explaining the
independence of $N$(\os) to metallicity.
For $T$=$3\!\times\!10^5$\,K (where \os\ peaks in CIE) and
$f$(\os)=0.22, the equation evaluates to  
$N$(\os)=$3\!\times\!10^{14}\,(v_{\rm cool}/100\,{\rm km s}^{-1})$,
implying a single cooling flow could explain the observed
galactic column of $\langle$log\,$N$(\os)$\rangle$=14.5 
if $v_{\rm cool}\!\approx\!100$\kms.
However, the kinematic complexity of galaxy halos makes such
single-zone solutions unlikely. More plausibly, the characteristic
value could be explained by a multi-zone cooling-flow scenario, where
each galactic sightline intersects a number of cooling regions.
If each region contributes an \os\ column of $\approx$10$^{13}$\sqcm,
the amount seen in the nearby LISM directions
\citep{SL06, Ba10} and predicted by conductive interface models
\citep{Bo90, GS10},  
then $\approx$30 regions are required along the kpc- to
tens-of-kpc scale sightlines through galactic halos to build up the
characteristic column.
In this multi-zone explanation, the constancy of $N$(\os) would be due 
not only to a similar \os\ column per interface, but also to a
similar number of interfaces intersected by each sight line.

The framework of the H02 model allows the cooling flows
to take several forms, including 
turbulent \citep{BF90, Sl93, Es06, KS10}, 
conductive \citep{Co79, BH87, Bo90, GS10}, 
or shocked \citep{DS96, IS04a, GS09}
interfaces between warm ($\sim$10$^4$\,K) interstellar/intergalactic 
clouds and surrounding hot ($\sim$10$^6$\,K) plasma.
Comparing the detailed predictions of these models 
to the \os\ observations is beyond the scope of this paper,
but there are evidently many theoretical reasons to expect
radiatively-cooling boundary layers to exist in the ISM and IGM.

H02 argued that the cooling-flow model is supported by
the \os\ log\,$N$-$b$ relation \citep[see also][]{Le06, Da06, Tr08, Gr09}. 
That relation is not presented here for the new compilation of \os\
absorbers, for two reasons. First, it is difficult to ensure
consistency and accuracy when comparing measurements of \os\ $b$-values, 
because they are 
(i) measured differently by different groups, 
(ii) difficult to measure in Galactic environments where
the \os\ lines are often saturated, and 
(iii) not necessarily resolved in spectra
of $\approx$20\kms\ resolution, as is the case with \fuse.
Second, it is unclear why $v_{\rm cool}$ should be
associated with $b$ or with the non-thermal broadening parameter
$b_{\rm nt}$, an implicit assumption 
when using the log $N$-$b$ relation to test the cooling-flow hypothesis.
Instead, here it is proposed that
$v_{\rm cool}$ could be identified with the velocity-centroid
offset $\delta v_0$ between \os\ and \hi\ 
(or between \os\ and a low-ion line such as \ion{C}{2}), because for a given
absorber, $\delta v_0$ represents a direct measurement of the relative
motion along the line-of-sight between the \os-bearing- and the low-ion
plasma phases.
Formally, $\delta v_0$ gives a lower limit on the absolute value
of $v_{\rm cool}$, since it only measures the line-of-sight component.
Unfortunately, measurements of $\delta v_0$ are not readily
available for the samples shown in Figures 1 and 2, but 
a search for a $N$(\os)-$\delta v_0$ relation
could form a future test of the cooling-flow model.

\section{Compilation of Intergalactic \os}

In Figure 3 the new compilation of intergalactic \os\ absorbers is presented.
It contains \nigm\ individual \os\ \emph{components} (often grouped
into systems) measured with line-profile fitting, and was formed by combining 
the low-$z$ sample of \citet[][hereafter T08]{Tr08}, 
the intermediate-$z$ samples of \citet[][hereafter R01 and R06]{Re01,
Re06}, and the high-$z$ samples of \citet[][hereafter BH05]{BH05} and
\citet[][hereafter F08]{Fo08a}.
The low-$z$ sample consists of 77 intervening and 34 proximate
components at a mean redshift $\langle z\rangle$=0.23, 
measured at $\approx$7\kms\ resolution with \hst/STIS/E140M and
a limiting column density (sensitivity) log\,$N_{\rm lim}\approx$13.0 (T08).
The intermediate-$z$ sample consists of 18 intervening and 5 proximate
components at $\langle z\rangle$=1.59,
measured at $\approx$10\kms\ resolution 
and log\,$N_{\rm lim}\approx$13.2 sensitivity with \hst/STIS/E230M (R01, R06).
The high-$z$ sample consists of 146 intervening components 
at $\langle z\rangle$=2.27 (BH05), 
and 48 proximate components at $\langle z\rangle$=2.36 (F08),
all measured at 6.6\kms\ resolution and log\,$N_{\rm lim}\approx$12.8
sensitivity with VLT/UVES.
All measurements were taken directly from the online version of the
papers cited (except the BH05 measurements, provided by J. Bergeron).
The intergalactic compilation is summarized in Table 2.

\begin{figure*}[!ht]
\epsscale{1.12}
\plotone{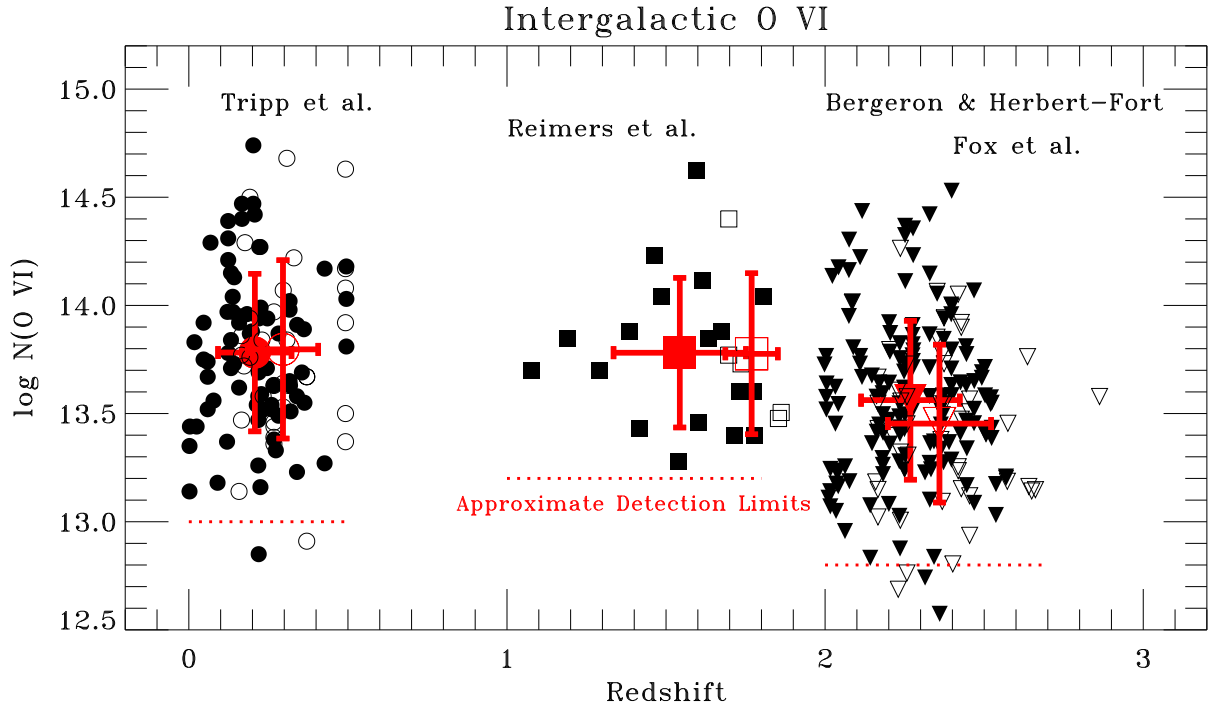}
\caption{Column density in {\it intergalactic} \os\ components as a
  function of redshift, using results from line-profile fitting.
  Shown are low-$z$ data from T08, intermediate-$z$ data from R01 and R06, 
  and high-$z$ data from BH05 and F08. 
  Proximate absorbers (within 5\,000\kms\ of the QSO redshift) are
  shown with open symbols. Red symbols with error bars show the mean
  and standard deviation of log\,$N$(\os) and $z$ in each
  sample. Approximate detection limits for each sample are shown with
  red dotted lines.}
\end{figure*}

Two interesting conclusions can be drawn from Figure 3. 
First, although there is significant scatter at all redshifts, 
the mean column density of intergalactic \os\ components evolves
only weakly over cosmic time.  
Second, at both low and high redshifts, there is no
difference in the mean \os\ component column density between
the intervening and proximate samples. These related results are now
discussed at length.
 
For intervening (non-proximate) \os\ components, the mean and standard
deviation of log\,$N$(\os) for components above a limiting
log\,$N$(\os)=13.2 changes from 
13.65$\pm$0.31 at $\langle z\rangle$=2.28 (BH05), to
13.78$\pm$0.35 at $\langle z\rangle$=1.54 (R01, R06), and
13.82$\pm$0.33 at $\langle z\rangle$=0.21 (T08), %*
where the cutoff at 13.2 (corresponding to a limiting rest-frame equivalent
width of 20\,m\AA) was used to correct for the small difference in
sensitivity between the samples. That amounts to a weak evolution 
in the mean log\,$N$(\os) of only 0.17\,dex (a factor of 1.5)
over 8.2\,Gyr of cosmic time
between $\langle z\rangle$=2.28 and $\langle z\rangle$=0.21 %* 
\citep[using the online cosmology calculator of][]{Wr06}.
This evolution is weak considering that cosmological simulations predict
a globally-averaged metallicity that increases from $\approx$0.01 solar at
$z$=3 to $\approx$0.2 solar at $z$=0 \citep{CO99}, i.e. by a factor of
$\approx$20 (although the local metallicity is highly dependent on
overdensity). 
Using the median rather than the mean log\,$N$(\os) does not
significantly change any of these results.
Kendall rank correlation tests were used to investigate whether 
log\,$N$(\os) and $z$ are statistically related.
A correlation coefficient of $-$0.17 was found at high significance,
indicating that the evolution on log\,$N$ is weak but real.
The same correlation coefficient ($-$0.17) was found 
when using the entire (intervening+proximate) sample.
The weak evolution in the mean \os\ column density
is highlighted in Figure 4, which compares the distribution of log\,$N$ 
of intervening \os\ components in the three redshift bins.

\begin{figure}[!ht]
\epsscale{1.15}
\plotone{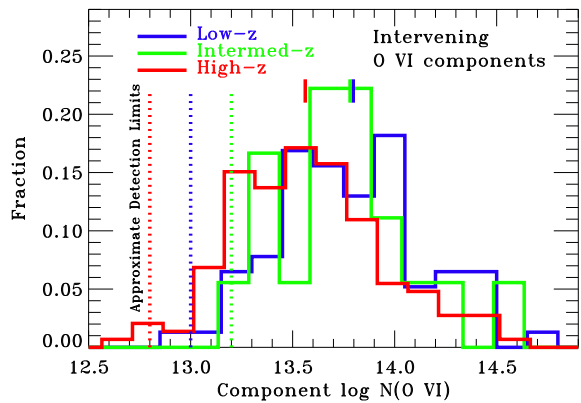}
\caption{Distribution of column density in intervening \os\ components 
  at low-$z$ (T08), intermediate-$z$ (R01, R06), 
  and high-$z$ (BH05) using a binsize of 0.15\,dex. The mean
  value of each distribution is shown with a color-coded tick
  mark. Dotted lines show the approximate detection limit
  of each sample.} 
\end{figure}

It is worth considering the potential impact of systematic effects on
the weak evolution of $N$(\os),
particularly because the intergalactic compilation is formed from
several heterogeneous datasets. 
The sensitivity limit of a given spectrum
(the limiting \os\ column density) is not straightforward to describe by
a single number, often varying from pixel to pixel, and so
the limits given on Figures 3 and 4 should be
considered as approximate, being set close to the column density of
the weakest component detected. At $z\!\ga\!2$, the increasing
density of the \lya\ forest leads to an increasing probability of a
given \os\ component being blended, which effectively reduces the
sensitivity to \os. When considering these effects, the \os\ column density
distributions at both low redshift \citep{DS08} and high redshift
(BH05) can be described by power laws. In this sense, the
observation of weak evolution in the mean log\,$N$(\os) does not
indicate the existence of an underlying characteristic value of 
intergalactic $N$(\os), as is the case for galactic \os\ absorbers,
but rather indicates that only part of the intergalactic
distribution is observable.  
Nonetheless, Figure 4 shows that in the range log\,$N>13.5$,
where all samples are expected to be complete, the distributions
of $N$(\os) in the three redshift bins show similar behavior,
overlapping up to log\,$N\!\approx\!14.5$. 
Therefore, accounting for sensitivity differences between
the datasets in the compilation reinforces
the conclusion that the \os\ populations are not strongly evolving.

If, for the low-redshift intergalactic sample,
results are taken from \citet{DS08} instead of T08, one finds
$\langle N$(\os)$\rangle$=13.66$\pm$0.45 instead of 13.78$\pm$0.36, %*
strengthening the result that the mean log\,$N$ of intergalactic \os\
evolves slowly with redshift. Since Danforth \& Shull present system-level
(integrated) rather than component-level column densities, 
this indicates that the weak evolution of log\,$N$(\os) remains true whether
one looks at components or systems. 

The other key result from Figure 3 is that
at both low and high redshifts, there is no
difference in the mean \os\ component column density between
the intervening and proximate samples 
(so long as truly intrinsic absorbers are excluded; \S3.3). 
At low redshift, the mean log\,$N$(\os) 
is 13.78$\pm$0.36 for intervening components and %* 
13.80$\pm$0.41 for proximate components (T08). %*
At high redshift, the mean log\,$N$(\os) 
is 13.56$\pm$0.37 for intervening components (BH05) and %*
13.45$\pm$0.36 for proximate components (F08). %*
Note that the \emph{incidence} d$N$/d$z$ of \os\ components does increase
with proximity to the quasar (within $\approx$2000\kms), at both
low-$z$ (T08) and high-$z$ (F08); our point here is that 
$\langle N$(\os)$\rangle$ does not.
This finding is important because, at both low and high redshift, the
intervening and proximate samples are measured in the \emph{same} set
of spectra, so the non-dependence of $N$(\os) on proximity cannot be simply a
sensitivity effect. 

Despite the weak evolution in the mean $N$(\os), there is an important
difference between the low-$z$ and high-$z$ populations, in that the
IGM \os\ components are, on average, almost twice as broad 
at low-$z$ than at high-$z$. This is shown in Figure 5,
where $b$(\os) is plotted against redshift for the intergalactic
compilation, with the low-$z$ and high-$z$ distributions compared on
the right. At low redshift, the mean and standard deviation of 
$b$(\os) for intervening components is 26$\pm$14\kms\ (T08). %*
At high redshift, the corresponding number is 14$\pm$7\kms\ (BH05). %*
This difference is not due to resolution, since the instrumental line widths
of each sample are much smaller than the mean \os\ line widths:
the instrumental FWHMs of 7\kms\ and 6.6\kms\ in the T08 and BH05 samples
correspond to $b$-values of 4.2\kms and 4.0\kms, respectively.

\begin{figure}[!ht]
\epsscale{1.15}
\plotone{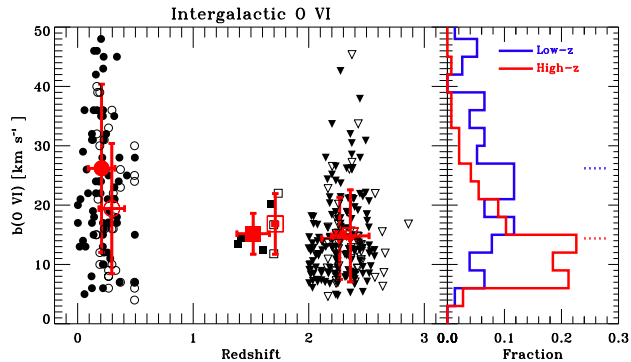}
\caption{Line-width ($b$-value) vs redshift for intergalactic \os\
  components, using low-$z$ data from T08, intermediate-$z$ data from
  R01 and R06, and high-$z$ data from BH05 and F08. Open symbols show
  proximate absorbers. The $b$-value distributions for the low-$z$ and
  high-$z$ intervening samples (the solid circles and solid triangles)
  are shown on the right-hand side. Dotted tick marks show the mean
  $b$-value of each distribution.} 
\end{figure}

\subsection{Discussion on Origin of Intergalactic \os}
The findings that $\langle N$(\os)$\rangle$ is insensitive to 
both redshift and quasar proximity are consistent; they show that
intergalactic \os\ absorbers are insensitive to the strength of the
local ionizing radiation field. 
This result has important implications if the IGM \os\ absorbers 
are photoionized by the EGB radiation, 
as is often argued \citep[e.g.][BH05]{Lv03a, TC08b, OD09},
because the production of \os\ by EGB photoionization (at any redshift)
requires an ionization parameter
$U\!\equiv\!n_\gamma/n_{\rm H}$ in the range $\approx$0.1--1 (see \S2).
Since $n_\gamma$ rises monotonically with $z$ \citep{HM96, HM01},
the non-variation of the mean $N$(\os) between $z$=0.2 and
$z$=2.3 can only be explained by photoionization if
$n_{\rm H}$ tracks $n_\gamma$, meaning that
the average gas density in the \os\ absorbers would have to
be $\approx$20 times higher at $z$=2.3 than at $z$=0.2. 
Since $n_{\rm H}$ scales as (1+$z$)$^3$ in an expanding Universe, such
a scaling may be physically reasonable (even expected). 
In turn, since the \os\ column density in any uniform-density 
absorber can be written as 
$N$(\os)=$n_{\rm H}d$(O/H)$f$(\os)=$n_\gamma d$(O/H)$f$(\os)/$U$,
where $d$ is the line-of-sight size of the cloud, 
%(O/H) is the oxygen abundance, and $f$(\os) is the \os\ ionization fraction,
the product $d$(O/H)$f$(\os) would have to 
decrease with redshift in exact proportion to the increase in
$n_\gamma$ in order for EGB photoionization to work, i.e. for it to
explain the non-variation of $N$(\os). This would require the high-$z$
absorbers to have smaller sizes, lower metallicities, or lower \os\
ionization fractions (or some combination of these three) than
their low-$z$ counterparts.

On the other hand, the H02 model, in which both galactic and
intergalactic \os\ absorbers are formed in radiatively-cooling regions
of initially-hot shock-heated plasma passing through the coronal regime,
offers a unified explanation for the non-evolution of $N$(\os) 
\emph{without any need for tunable parameters or for photoionization}.
Explaining the mean intergalactic
log\,$N$(\os)$\approx$13.7 by the cooling-flow model would require 
a single region with $v_{\rm cool}$(IGM)$\approx$20\kms, or a multi-zone
arrangement such as $\approx$5 interfaces each contributing
$\approx$10$^{13}$\sqcm. The idea that the plasma traced by
intergalactic \os\ is cooling rather than photoionized is further
supported by two model-independent kinematic observations: 

(i) the median line-width ($b$-value) is higher for 
\os\ than for other (photoionized) metal lines observed in the IGM. 
In their high-$z$ intervening samples,
BH05 and \citet{Si02} report a median $b$(\os) of
13\kms\ and 16\kms, respectively,
whereas significantly smaller median $b$-values are seen for
\cf\ \citep[$\langle b\rangle\!\approx\!9$\kms;][]{DO10} and  
\nf\ \citep[$\langle b\rangle\!\approx\!6$\kms;][]{FR09}.
At low-$z$, the IGM sample of \citet{DS08} has a median $b$(\os) of
27\kms (the T08 sample has 24\kms), again larger than the median
$b$-values for \nf\ (17\kms) and \cf\ (19\kms) observed in the same
redshift range \citep{DS08}.

(ii) significant velocity-centroid offsets (up to 20\kms)
exist between the \os, \hi, and \cf\ components in a large fraction
of intergalactic \os\ absorbers observed at low- (T08), intermediate-
(R01) and high- \citep[F08,][]{Si02, FR09} redshift. For the T08
sample, only 28 of 77 intervening \os\ components (36\%) are aligned
with \hi\ components. 

For (the majority of) IGM absorbers with $b$-value differences and/or
velocity-centroid offsets between \os\ and other species, multi-phase
solutions are required, and single-phase photoionization models cannot
be used; their use will give physically irrelevant results.
Even in (the minority of) IGM absorbers with aligned \os\ and \hi\
components, there is no guarantee of co-spatiality of the two ions;
simulations show cases of coincident absorption arising from spatially 
distinct regions of gas \citep{OD09}. Therefore the observation that
\os\ and \hi\ components are occasionally aligned should not be taken
as proof of single-phase photoionization. Such alignments can easily
be explained by interface theories \citep[e.g.][]{BH87}. 
Finally, the high-$z$ mean \os\ $b$-value of 14\kms\ implies a
plasma temperature log\,$T\!<$5.27, close to the temperature at which
\os\ peaks in CIE, and easily explainable by non-equilibrium CI
models \citep[e.g.][]{GS07}, which indicate that \os\ can exist at
temperatures below $10^5$\,K if the metallicity is high enough. 

\section{Summary and Conclusions}
An extensive heterogeneous compilation of \ntot\ \os\ absorbers observed at
high-resolution (instrumental FWHM$\le$20\kms) has been presented,
covering the Local ISM, the disk and halo of the Milky Way, HVCs, the
SMC, the LMC, starburst galaxies, the intergalactic medium from $z$=0
to $z\!\approx\!2.5$, DLAs at $z$=2--3, and GRB host
galaxies at $z$=2--4, using results drawn from the literature. 
This compilation is divided into a galactic sample 
[defined by log\,$N$(\hi)$>$17, and usually log\,$N$(\hi)$\ga$20] 
consisting of \ngal\ \os\ measurements in galaxies with
metallicities from $\approx$0.01 solar to solar, and an intergalactic
sample [log\,$N$(\hi)$<$17] consisting of \nigm\ \os\ components
covering redshifts from $z$=0 to $z\!\approx\!2.5$. The key
observational findings are: 

\begin{enumerate} 
\item For \os\ measured in extended sight lines 
through {\bf galactic} halos, the mean value of log\,$N$(\os) 
is surprisingly insensitive to metallicity (and mass), 
with the Milky Way halo, the LMC, the SMC, starburst galaxies, and
DLAs at $z$=2--3 all showing a mean log\,$N$(\os) within 0.4\,dex
of 14.5, even though they span $\approx$2\,dex in [O/H].
While this characteristic value has been noticed before in the local
Universe, the new result is that it
applies down to the DLA regime at $-$1.6$\la$[O/H]$\la$--0.6, though
there is a suggestion from the lowest-metallicity DLAs that $\langle
N$(\os)$\rangle$ falls off at [O/H]$<$--1.6. 

\item For {\bf intergalactic} \os,
there is surprisingly little evolution in the mean \os\ component column
density over cosmic time. Over the 8.2\,Gyr interval
between $z$=2.28 and $z$=0.21, the mean log\,$N$ %*
of intervening \os\ components with log\,$N$(\os)$>$13.2
(corresponding to rest equivalent widths $>$20\,m\AA)
increases by only 0.17\,dex, a factor of 1.5. %*
The distributions of $N$(\os) at log\,$N\!>\!13.5$,
where the samples are complete, are shown to be insensitive to redshift.
Furthermore, at both low and high redshifts, there is no
difference in the mean log\,$N$(\os) between
the intervening and proximate samples 
(so long as truly intrinsic absorbers are excluded). 
The insensitivity of log\,$N$(\os) to redshift and quasar proximity
indicates an insensitivity to the strength of the ionizing 
radiation field, which (using the extragalactic background) is a
factor of $\approx$20 higher at $z$=2.28 than at $z$=0.21.

\item Intergalactic \os\ components are, on average, almost twice as broad
  at low-$z$ than at high-$z$, with the mean $b$(\os) rising from
  14\kms\ at $\langle z\rangle$=2.28 to 26~\kms\ at 
  $\langle z\rangle$=0.21. %*
\end{enumerate}

The observation of a ``characteristic'' \os\ column density in many
diverse galactic halos covering a range of mass and metallicity, 
is suggestive of a common origin or regulation mechanism. 
One such potential origin is the cooling-flow model of H02.
In this model both galactic and intergalactic \os\
absorbers trace regions of initially-hot shock-heated plasma that are
now radiatively cooling through coronal temperatures.
The key advantage of this model is that it naturally
explains the insensitivity of $N$(\os) to metallicity and redshift,
though it is unclear how it could explain the evolution in
IGM $b$-values (result 3).
The general framework of the cooling-flow theory allows the regions to
take several forms, including conductive, turbulent, or shocked
interfaces between warm ($\sim$10$^4$\,K) clouds and hot
($\sim$10$^6$\,K) plasma. However, many such regions would have to be
intersected by a typical galaxy-halo sightline to build up the
characteristic galactic $N$(\os), which is $\approx$30
times larger the column predicted in a single conductive interface.

The idea that much of the galactic \os\ arises in
coronal-temperature boundary layers is well-known and
well-supported (see references in \S3.1). The idea that
\emph{intergalactic} \os\ absorbers are also produced in such boundary
layers, instead of by photoionization, is more controversial.
The newly-demonstrated insensitivity of the mean intergalactic
$N$(\os) to $z$ can only be explained by photoionization if the gas
density tracks the ionizing photon density, which would require the
high-$z$ \os\ absorbers to have smaller sizes, lower metallicities,
and/or lower ionization fractions at high-$z$ than the low-$z$ absorbers.
While this is plausible (even expected), the kinematics of
most intergalactic \os\ absorbers observed at low and high redshift
(specifically the $b$-value differences and velocity-centroid offsets
observed between \os, \cf, and \hi) rule out single-phase
photoionization, and require multi-phase models such as the
cooling-flow scenario.

Intergalactic \os\ absorbers are often discussed in the context
of the elusive warm-hot intergalactic medium (WHIM),
predicted by cosmological simulations to contain 
a substantial fraction of the present-day baryons
\citep{CO99, CO06, Ce01, Da01, FB01, Ka05, CF06}.
The results presented here support the view that although \os\
absorbers do not 
directly trace the (hotter) bulk of the WHIM \citep{OD09, TG10, Sm10}, 
they do trace the boundary layers where the WHIM interfaces with
cooler, metal-enriched regions.\\ 

{\it Acknowledgments.}
The author acknowledges valuable conversations with
Blair Savage, Benjamin Oppenheimer, Linda Smith, and Bart Wakker,
and is grateful to the anonymous referee for an insightful report.
He thanks Jacqueline Bergeron for providing her \os\ sample in
electronic form.
He appreciates the support of an ESO Fellowship, and a Marie Curie
Intra-European Fellowship (contract MEIF-CT-2005-023720) 
awarded to investigate \os. %in 2006 and 2007.
This paper has made considerable use of the NASA Astrophysics
Data System Abstract Service.

\begin{deluxetable*}{lllll ll}
\tabletypesize{\footnotesize}
\tablecaption{Compilation of Galactic \os\ absorbers [log\,$N$(\hi)$>$17]}
\tablehead{Location & Number\tm{a} & $\langle z \rangle$ & $\langle$[O/H]$\rangle$\tm{b} & $\langle$log\,$N$(\os)$\rangle$\tm{c} & Instr. & Reference}
\startdata
                     {\it Local ISM}\tm{d} &  24    & 0 & \nodata & 13.09$\pm$0.30       & \fuse\ & Savage \& Lehner (2006)\\
   Galactic disk ($|b|\!<\!10\degr$)\tm{e} & 131                & 0 & $\sim$0.0  & 14.15$\pm$0.35           & \fuse\ & Bowen et al. (2008)\\
   Galactic halo ($|b|\!>\!10\degr$)\tm{f} &  91              & 0 & $\sim$--0.25 & 14.36$\pm$0.18          & \fuse\ & Savage et al. (2003)\\
   {\it High-velocity clouds (HVCs)}\tm{g} &  84              & 0 & $\sim$--0.25 & 13.95$\pm$0.34         & \fuse\ & Sembach et al. (2003)\\
                                 SMC\tm{h} &  18    & 0.00053 & $-$0.66$\pm$0.10 & 14.58$\pm$0.17          & \fuse\ & Hoopes et al. (2002)\\
                                 LMC\tm{i} &  70    & 0.00094 & $-$0.34$\pm$0.06 & 14.24$\pm$0.19          & \fuse\ & Pathak et al. (2010)\\
                  Starburst galaxies\tm{j} &  12 & 0.0128 & $-$0.36$\pm$0.47 & 14.92$\pm$0.34          & \fuse\ & Grimes et al. (2009)\\
                                DLAs\tm{k} &  12 & 2.56 & $-$1.62$\pm$0.61 & 14.47$\pm$0.54              & UVES & Fox et al. (2007a)\\
           ... high-met. ([Z/H]$>$--1.63)  & (6) & 2.46 & $-$1.18$\pm$0.50 & 14.79$\pm$0.49                  & $\prime\prime$ & $\prime\prime$\\
            ... low-met. ([Z/H]$<$--1.63)  & (6) & 2.65 & $-$2.06$\pm$0.34 & 14.16$\pm$0.42                  & $\prime\prime$ & $\prime\prime$\\
              GRB host galaxies\tm{m} &   5 & 2.72 & \nodata & $>$14.81$\pm$0.22                & UVES & Fox et al. (2008b)
\enddata
\tablecomments{Entries in italics denote partial (rather than extended) galaxy-halo sightlines.}
\tn{a}{Number of sightlines in sample, except for HVCs, where number of individual components is given. Numbers for sub-samples given in parentheses.\\}
\tn{b}{Mean gas-phase metallicity (logarithmic scale relative to solar).\\}
\tn{c}{Mean and standard deviation of logarithmic \os\ column density in \sqcm.\\}
\tn{d}{~Local ISM results given here are for \os\ detections only; 15 non-detections and 7 cases with possible stellar contamination are ignored. Columns measured along short ($\approx$40--200\,pc) sightlines.\\}
\tn{e}{Galactic disk results given here are for \os\ detections only and are derived from Voigt-profile fits; 17 non-detections are ignored. Solar metallicity assumed.\\}
\tn{f}{Galactic halo columns integrated over $-100<v_{\rm LSR}\!<\!100$\kms. Results given here are for \os\ detections only; 11 non-detections are ignored. $\approx$0.5 solar metallicity assumed.\\}
\tn{g}{HVCs are individual clouds with $|v_{\rm LSR}|\!>\!100$\kms. The 84 HVCs were measured along 100 sightlines. $\approx$0.5 solar metallicity assumed.\\}
\tn{h}{SMC columns integrated over $v_{\rm LSR}\approx$50--260\kms. [O/H]$_{\rm SMC}$ from \citet{RD92}.\\}
\tn{i}{~LMC columns integrated over $v_{\rm LSR}\approx$160--360\kms. [O/H]$_{\rm LMC}$ from \citet{RD92}.\\}
\tn{j}{Starburst \os\ columns integrated over full velocity ranges observed, typically $\approx$200--300\kms; 4 non-detections are ignored. Metallicities derived from nebular emission lines.\\}
\tn{k}{DLA results are for 12 DLAs with \os\ detections; 23 cases with blended \os\ are ignored. DLA metallicities shown are [Z/H] where Z=Zn, S, or Si.\\}
\tn{m}{GRB host galaxy \os\ columns integrated over full velocity ranges observed, typically $\approx$200\kms\ around the burst redshift. All individual GRB \os\ columns are lower limits (due to saturation), so the mean is also a lower limit. GRB host galaxy metallicities are not well-characterized and are not shown.\\}
\end{deluxetable*}

\begin{deluxetable*}{lllll ll}
\tablewidth{0pt}
\tabletypesize{\footnotesize}
\tablecaption{Compilation of Intergalactic \os\ components [log\,$N$(\hi)$<$17]}
\tablehead{Sample & Number\tm{a} & $\langle z \rangle$ &$\langle$log\,$N$(\os)$\rangle$\tm{b} & $\langle b$(\os)$\rangle$\tm{c} & Instr. & Reference}
\startdata
                  Low-$z$ intervening &  77 & 0.21 & 13.78$\pm$0.36 & 26.2$\pm$14.2                    & STIS & Tripp et al. (2008)\\
                ... log $N>13.2$ only & (73) & 0.21 & 13.82$\pm$0.33 & 26.8$\pm$14.2                         & STIS & $\prime\prime$\\
                    Low-$z$ proximate &  34 & 0.30 & 13.80$\pm$0.41 & 19.4$\pm$11.0                         & STIS & $\prime\prime$\\
                ... log $N>13.2$ only & (32) & 0.30 & 13.85$\pm$0.37 & 19.9$\pm$11.1                         & STIS & $\prime\prime$\\
   Intermediate-$z$ intervening\tm{d} &  18 & 1.54 & 13.78$\pm$0.35 & 15.2$\pm$3.4            & STIS & Reimers et al. (2001, 2006)\\
     Intermediate-$z$ proximate\tm{d} &   5 & 1.77 & 13.78$\pm$0.37 & 16.8$\pm$5.1                         & STIS & $\prime\prime$\\
                 High-$z$ intervening & 146 & 2.27 & 13.56$\pm$0.37 & 14.4$\pm$6.9        & UVES & Bergeron \& Herbert-Fort (2005)\\
                ... log $N>13.2$ only & (125) & 2.28 & 13.65$\pm$0.31 & 15.1$\pm$7.1                         & UVES & $\prime\prime$\\
                   High-$z$ proximate &  48 & 2.36 & 13.45$\pm$0.36 & 14.8$\pm$7.8                     & UVES & Fox et al. (2008a)\\
                ... log $N>13.2$ only & (33) & 2.35 & 13.64$\pm$0.26 & 17.2$\pm$8.2                           & UVES & $\prime\prime$
\enddata
\tablecomments{\emph{Intervening} and \emph{proximate} refer here to components with velocity offsets from the background QSO of $>$5000 and $<$5000\kms, respectively. Quasar-intrinsic absorbers are excluded.\\}
\tn{a}{Sample size: number of components. Numbers for sub-samples given in parentheses.\\}
\tn{b}{Mean and standard deviation of logarithmic \os\ column density in \sqcm.\\}
\tn{c}{Mean and standard deviation of \os\ $b$-value in \kms.\\}
\tn{d}{All intermediate-$z$ absorbers have log\,$N>13.2$, so no sub-sample is needed.\\}
\end{deluxetable*}

\end{document}